\RequirePackage{fix-cm}
\documentclass[twocolumn,epjc3]{svjour3}
\smartqed  
\usepackage{amsmath}
\usepackage{amssymb}
\usepackage{amssymb,bbold}

\RequirePackage{graphicx}
\RequirePackage[colorlinks,citecolor=blue,urlcolor=blue,linkcolor=blue]{hyperref}
\RequirePackage[numbers,sort&compress]{natbib}

\journalname{Eur. Phys. J. C}
\begin{document}

\title{Comment on ``Comment on Linear confinement of a scalar particle in a G\"{o}del-type space-time''
}


\author{Francisco A. Cruz Neto\thanksref{addr1}\and Camila C. Soares\thanksref{addr1} \and Luis B. Castro\thanksref{e1,addr1,addr2} 
}

\thankstext{e1}{e-mail: luis.castro@ufma.br, luis.castro@pq.cnpq.br}


\institute{Departamento de F\'{\i}sica, Universidade Federal do Maranh\~{a}o, Campus Universit\'{a}rio do Bacanga, 65080-805, S\~{a}o Lu\'{\i}s, MA, Brazil.
\label{addr1} \and
Departamento de F\'{\i}sica e Qu\'{\i}mica, Universidade Estadual Paulista (UNESP), Campus de Guaratin\-gue\-t\'{a}, 12516-410, Guaratinguet\'{a}, SP, Brazil.\label{addr2}
}

\date{Received: date / Accepted: date}

\maketitle

\begin{abstract}
We show that from an appropriate manipulation of the biconfluent Heun differential equation can obtain the correct expression for the energy eigenvalues for the Klein-Gordon equation without potential in the background of Som-Raychaudhuri space-time with a cosmic string as a case particular ($k_{L}=0$) of [Vit\'{o}ria et al. Eur. Phys. J. C (2018) 78:44], in opposition what was stated in a recent paper published in this journal [F. Ahmed, Eur. Phys. J. C (2019) 79:682]. 

\end{abstract}


In a recent paper in this Journal, F. Ahmed \cite{EPJC79:682:2019} has pointed out an incorrect expression in \cite{EPJC78:44:2018}. The author has solved the Klein-Gordon equation without interaction in the Som-Raychaudhuri space-time with the cosmic string using the Nikiforov-Uvarov method and obtained the energy eigenvalues and the corresponding eigenfunctions of the system. The problem solved in \cite{EPJC79:682:2019} is a special case associates to $k_{L}=0$ in Ref. \cite{EPJC78:44:2018}. F. Ahmed has claimed that to obtain the correct result, one should start from Eq. (4) but not from the condition (12), which is obtained from the biconfluent Heun equation (10). The purpose of this comment is to show that following an appropriate manipulation of the biconfluent Heun equation, the correct energy eigenvalues for $k_{L}=0$ can be obtained as a particular case of \cite{EPJC78:44:2018}, as opposed to what was advertised in Ref. \cite{EPJC79:682:2019}.

The Klein-Gordon equation in the Som-Raychaudhuri space-time with a linear scalar potential is given by
\begin{equation}\label{kge}
\begin{split}
-\frac{\partial^{2}\Psi}{\partial t^{2}}+\frac{1}{r}\frac{\partial}{\partial r}\left(r\frac{\partial\Psi}{\partial r}\right)+\left( \frac{1}{\alpha r}\frac{\partial}{\partial \phi}-\Omega r\frac{\partial}{\partial t} \right)^{2}\Psi\\
+\frac{\partial^{2}\Psi}{\partial z^{2}}=\left( M+k_{L}r \right)^{2}\Psi\,.
\end{split}
\end{equation}
\noindent Considering the solution in the form 
\begin{equation}\label{sol1}
\Psi(t,r,\phi,z)=\mathrm{e}^{i(-Et+l\phi+kz)}\psi(r)\,,
\end{equation}
\noindent and by introducing the new variable and parameters
\begin{equation}\label{newvar}
x=\sqrt{\omega}r\,,
\end{equation}
\begin{equation}\label{omega}
\omega=\sqrt{\Omega^{2}E^{2}+k_{L}^{2}}\,,
\end{equation}
\begin{equation}\label{lambda}
\lambda=E^{2}-M^{2}-k^{2}-\frac{2\Omega El}{\alpha}
\end{equation}
\begin{equation}
\theta=\frac{2Mk_{L}}{\omega^{3/2}}\,,
\end{equation}
\begin{equation}\label{beta}
\beta=\frac{\lambda}{\omega}\,,
\end{equation}
\noindent one finds that equation (\ref{kge}) becomes
\begin{equation}\label{eq1}
\psi^{\prime\prime}+\frac{1}{x}\psi^{\prime}+\left( \beta-x^{2}-\frac{l^{2}}{\alpha^{2}x^{2}}-\theta x \right)\psi=0\,.
\end{equation}
\noindent The solution for (\ref{eq1}) can be expressed as
\begin{equation}\label{sol2}
\psi(x)=x^{\frac{|l|}{\alpha}}\mathrm{e}^{-\frac{x^{2}}{2}}\mathrm{e}^{-\frac{\theta}{2}x}H(x)\,,
\end{equation}
\noindent where $H(x)$ can be expressed as a solution of the biconfluent Heun differential equation \cite{JPA19:3527:1986,RONVEAUX1995,PRC86:052201:2012,EPJC78:494:2018}
\begin{equation}\label{bche}
\begin{split}
H^{\prime\prime}+\left( \frac{1+\frac{2|l|}{\alpha}}{x}-\theta-2x \right)H^{\prime}\\
+\left( \tau-2-\frac{2|l|}{\alpha}- \frac{\sigma}{x}\right)H=0\,,
\end{split}
\end{equation}
\noindent with 
\begin{equation}
\tau=\beta+\frac{\theta^{2}}{4}\,,
\end{equation}
\noindent and
\begin{equation}
\sigma=\frac{\theta}{2}\left( 1+\frac{2|l|}{\alpha} \right)\,.
\end{equation}
\noindent The differential equation (\ref{bche}) has a regular singularity at $x=0$ and an irregular singularity at $x=\infty$. The regular solution at the origin is given by
\begin{equation}\label{sol3}
H(2|l|/\alpha,\theta,\tau,0;x)=\sum_{j=0}^{\infty}\frac{1}{\Gamma(1+j)}
\frac{A_{j}}{j!}x^{j}\,,
\end{equation}
\noindent where $\Gamma(z)$ is the gamma function, $A_{0}=1$, $A_{1}=\sigma$ and the remaining coefficients for $\theta\neq 0$ satisfy the recurrence relation,
\begin{equation}\label{recu}
\begin{split}
A_{j+2}= & \frac{\theta}{2}\left( 2j+3+\frac{2|l|}{\alpha} \right)A_{j+1}\\
&-\left(j+1\right)\left( j+\frac{2|l|}{\alpha}+1 \right)\left(\Delta-2j\right)A_{j}\,,
\end{split}
\end{equation}
\noindent where $\Delta=\tau-\frac{2|l|}{\alpha}-2$. From the recurrence (\ref{recu}), the solution $H$ becomes a polynomial of degree $n$ if and only if $\Delta=2n$ ($n=0,1,\ldots$) and $A_{n+1}=0$. 

On the other hand, if $\theta=0$, the solution $H$ becomes a polynomial of degree $n$ if and only if 
\begin{equation}\label{qc2}
\Delta=4n \quad (n=0,1,\ldots)\,.
\end{equation}
\noindent In fact, when $\Delta=4n$, one has \cite{RONVEAUX1995}
\begin{equation}\label{bchL}
H(2|l|/\alpha,0,2|l|/\alpha+2(1+2n),0;x)\propto L^{(|l|/\alpha)}_{n}(x^{2})\,,
\end{equation} 
\noindent where $L^{(\delta)}_{n}(x)$ denotes the generalized Laguerre polynomial, a polynomial of degree $n$ with $n$ distinct positive zeros in the range $[0,\infty)$. This last result is very important to obtain the correct energy eigenvalues for the particular case $k_{L}=0$ ($\theta=0$).

The especial case $k_{L}=0$ ($\theta=0$) was studied in \cite{EPJC79:682:2019}. The author of \cite{EPJC79:682:2019} concluded that the correct expression of energy cannot be obtained as a particular case from the biconfluent Heun differential equation, but this statement is false. Considering $k_{L}=0$ ($\theta=0$) and using the correct condition (\ref{qc2}), we obtain
\begin{equation}\label{ener1}
\beta-\frac{2|l|}{\alpha}-2=4n\,.
\end{equation}  
\noindent Substituting (\ref{omega}), (\ref{lambda}) and (\ref{beta}) into (\ref{ener1}), we obtain the spectrum for $k_{L}=0$ as (for $E\Omega>0$)
\begin{equation}\label{spectrum}
\begin{split}
E=&\left(2n+1+\frac{|l|}{\alpha}+\frac{l}{\alpha}\right)\Omega\\
&+\sqrt{\left( 2n+1+\frac{|l|}{\alpha}+\frac{l}{\alpha} \right)^{2}\Omega^{2}+M^{2}+k^{2}}\,,
\end{split}
\end{equation}
\noindent and the solution becomes
\begin{equation}\label{solution}
\psi(x)=N_{n}x^{\frac{|l|}{\alpha}}\mathrm{e}^{-\frac{x^{2}}{2}}L_{n}^{(|l|/\alpha)}(x^{2})\,,
\end{equation}
\noindent where $N_{n}$ is a normalization constant. The expression (\ref{spectrum}) is the same to the result obtained in Ref. \cite{EPJC74:2935:2014}.

In summary, we showed that the correct expression for the energy eigenvalues for the case $k_{L}=0$ of Ref. \cite{EPJC78:44:2018} can be obtained from an appropriate manipulation of the biconfluent Heun differential equation, in opposition what was concluded in \cite{EPJC79:682:2019}. The results obtained in this comment are consistent with those found in \cite{EPJC74:2935:2014}.

\begin{acknowledgements}
This work was supported in part by means of funds provided by CNPq, Brazil, Grant No. 307932/2017-6 (PQ) and No. 422755/2018-4 (UNIVERSAL), S\~{a}o Paulo Research Foundation (FAPESP), Grant No. 2018/20577-4, FAPEMA, Brazil, Grant No. UNIVERSAL-01220/18 and CAPES, Brazil.
\end{acknowledgements}


\end{document}